\documentclass[runningheads,a4paper]{llncs}

\usepackage{amssymb}
\usepackage{graphicx}

\usepackage{url}

\usepackage{textcomp}

\usepackage{array}

\urldef{\mailsa}\path|sagatov@ya.ru, amskh@yandex.ru|    
\newcommand{\keywords}[1]{\par\addvspace\baselineskip
\noindent\keywordname\enspace\ignorespaces#1}

\begin{document}

\mainmatter

\title{Duplication of Key Frames of Video Streams\\in Wireless Networks}

\titlerunning{Duplication of Key Frames of Video Streams in Wireless Networks}

\author{Evgeny S. Sagatov \and Andrei M. Sukhov}

\authorrunning{Duplication of Key Frames of Video Streams in Wireless Networks}

\institute{Samara State Aerospace University,\\
Moskovskoe sh. 34, 443086 Samara, Russia\\
\mailsa\\}

\toctitle{Duplication of Key Frames of Video Streams in Wireless Networks}
\tocauthor{Evgeny S. Sagatov, Andrei M. Sukhov}
\maketitle

\begin{abstract}
In this paper technological solutions for improving the quality of video transfer along wireless networks are investigated. Tools have been developed to allow packets to be duplicated with key frames data. In the paper we tested video streams with duplication of all frames, with duplication of key frames, and without duplication. The experiments showed that the best results are obtained by duplication of packages which contain key frames. The paper also provides an overview of the coefficients describing the dependence of video quality on packet loss and delay variation (network jitter).
\keywords{video quality, wireless network, RTP stream, IPTV}
\end{abstract}

\section{Introduction}
Human mobility assumes an instantaneous connection to the Internet from anywhere and at any time. Technological solutions for mobility can be provided by the wireless networks of different standards that surround modern man everywhere. In the big city we are always in an area of several mobile networks of different generations: Wi-Fi, 3G, and WiMAX. These services are very often available completely free in entertainment centres, shops, restaurants, and hotels for promotional purposes.

The absence of wires and of being tied to a fixed place makes the mobile Internet incredibly popular, and it is growing exponentially year by year~\cite{cisco}. Internet channels and technologies on such networks allow most of the services to be used easily. But mobility imposes restrictions on last generation applications such as real-time video, networked multimedia, Internet broadcasting, and so on. Their implementation is limited by qualitative parameters of wireless networks: a significant percentage packet loss and high values of packet delay variation. Details of the types of quality network connection for video transfer are presented in~\cite{claypool,calyam}.

Earlier in~\cite{sagatov} the problem of adaptation of modern encoding algorithms and video transfer along wireless networks such as 3G, Wi-Fi, and WiMAX~\cite{haghani}, as well as in other networks with poor quality characteristics, was addressed. An attempt was made to find and to compare the numerical correlation of video quality with the network settings. This dependence is described by a simple model that allowed us to use the numerical value of the coefficients to compare the quality of the video and determine the most significant factors, as well as compare the different codecs. In this paper we take into account differences between the distortions which damage the key and intermediate frames.

In this study we implement a research program outlined in~\cite{sagatov}, where several ways to improve the video quality in wireless networks were suggested. We will analyse the effect of packet duplication and adding redundant information to key frames on the video quality.

This article provides an overview that compares the quality characteristics of the major wireless networks such as Wi-Fi, 3G, and WiMAX. The effect of duplication of RTP packets stream on the video quality is also investigated. Additionally, we have carried out experiments on the 802.16e standard mobile WiMAX from the operator Yota (S. Peterburg) and analysed data for the WMV codec because the last version of VirtualDub can identify key frames in the corresponding video stream.

\section{Modelling}
Connection quality deteriorates~\cite{hei,won} depending on the characteristics of a network connection when video is transmitting over a network. The transmitted video quality is measured on a scale called mean opinion score (MOS)~\cite{itu} and can be described by a universal function $Q_{MOS}(p,j,D,B)$ depending on the network parameters \cite{bradner}:

\begin{itemize}
	\renewcommand\labelitemi{}
	\item $p$ -- percentage packet loss, \%;
	\item $j$ -- network jitter (delay variation $D$) at the time of error, seconds;
	\item $Q_{MOS}$ --  quality of the received video, on a scale of one to five;
	\item $B$ -- available bandwidth, Mbps.
\end{itemize}

This function can be expanded in a power series, and taking into account the small values of network variables we can restrict ourselves to linear terms.

In~\cite{sukhov} it was shown that for a video stream with a fixed bitrate it is sufficient to consider only the expansion terms describing the linear relationship between two variables, packet loss and network jitter:

\begin{equation}
	\label{eq:model}
	Q_{MOS}=Q_{ideal}-\alpha p- \beta j,	
\end{equation}

\begin{itemize}
	\renewcommand\labelitemi{}
	\item $Q_{ideal}$ -- maximum video quality for the codec, on a scale of zero to five;
	\item $\alpha, \beta$ -- model coefficients to be determined experimentally.
\end{itemize}

For research a single video sequence has been selected which was encoded with MPEG-4 (DivX), MPEG-2, and Windows Media Video 9 with a constant bitrate of 256 kbps. This sequence was used to test the video. This can be found on the global network at site~\cite{traces}.

\section{Planning of Experiment}
In order to calculate coefficient values from Equation~(\ref{eq:model}) we developed and conducted several experiments~\cite{sagatov}. Video files encoded as MPEG-4 (DivX), MPEG-2, and Windows Media Video 9 were transferred by RTP stream with VideoLan VLC~\cite{videolan} on a laptop connected through a wireless network using Wi-Fi, WiMAX, or 3G standards. We recorded incoming video on a notebook to the file using VideoLan VLC and simultaneously captured network traffic using a Wireshark sniffer~\cite{wireshark}. Thus, we can determine the video quality on a MOS scale for video and network connection settings obtained using network traces. We used the tools VirtualDub~\cite{virtualdub} and Avisynth~\cite{avisynth} to help with the analysis of video and VQMT~\cite{vatolin} to find $Q_{ideal}$.

All video files and network traffic traces recorded during the experiments are published on the ``Internet TV" website~\cite{traces}.

For the experiment one video fragment with different types of film: static, with slow motion, fast motion, and with a brightness change, was prepared. Then the video sequence was encoded using the MPEG-4 (DivX), MPEG-2, and Windows Media Video 9 codecs. In this fragment the following video settings are established:

\begin{itemize}
	\item video resolution: 320 x 240 pixels;
	\item frame rate: 24 fps (frames per second);
	\item bitrate: 256 Kbps (Kilobits per second);
	\item quality: maximal one.
\end{itemize}

For the experiments we used the network segments of the following operators: Megafon Samara (3G), Beeline Samara (3G), Metromaks Samara (fixed WiMAX), and Yota St. Petersburg (mobile WiMAX).

3G network experiments using the operators Megafon and Beeline were performed on UMTS standard equipment, which is commonly exploited by these operators.

\section{Model coefficients}
As a result of the experiments we obtained values of the coefficients for MPEG-2, MPEG-4 (DivX), and WMV9 codecs on Wi-Fi, 3G, and WiMAX networks, which are summarized in Tables 1--3. These coefficients are presented separately with errors in key and usual frames and sorted by different codecs and wireless networks standards. The following notations are used:

\begin{itemize}
	\renewcommand\labelitemi{}
	\item $\alpha^k$ -- the coefficient for the packet loss in key frames, whose values are expressed as a percentage;
	\item $\beta^k$ -- the coefficient for the network jitter, measured in seconds;
	\item $\alpha^w$ and $\beta^w$ -- coefficients of the model for the video segments without packet loss of key frames.
\end{itemize}

\begin{table}[!ht]
	\caption{\label{tablewifi}The values of the model coefficients for MPEG-2, MPEG-4 (DivX), and WMV9 codecs in a Wi-Fi network}
	\begin{center}
		\begin{tabular}{c@{\quad}c@{\quad}c@{\quad}c@{\quad}c@{\quad}c@{\quad}c}
		\hline\noalign{\smallskip}
		\textnumero & Codec &	$Q_{ideal}$ &	$\alpha^k$ & $\beta^k$  &	 $\alpha^w$ & $\beta^w$\\
		[2pt]
		\hline
		\noalign{\smallskip}
		1 &	MPEG-2 & 4.2$\pm$0.2 &	0.11$\pm$0.03	& 15$\pm$4 &	0.06$\pm$0.02 &	10$\pm$4 \\
		2 & DivX  & 4.7$\pm$0.2 &	0.25$\pm$0.05	& 15$\pm$5 &	0.17$\pm$0.02 &	10$\pm$3 \\ 
		3 & WMV9  & 4.7$\pm$0.2 &	0.25$\pm$0.11	& 20$\pm$8 &	0.16$\pm$0.6 &	10$\pm$3 \\
		[2pt]
		\hline
		\end{tabular}
	\end{center}
\end{table}

\begin{table}[!ht]
	\caption{\label{table3g}The values of the model coefficients for MPEG-2, MPEG-4 (DivX), and WMV9 codecs in a 3G network}
	\begin{center}
		\begin{tabular}{c@{\quad}c@{\quad}c@{\quad}c@{\quad}c@{\quad}c@{\quad}c}
		\hline%\noalign{\smallskip}
		\textnumero & Codec &	$Q_{ideal}$ &	$\alpha^k$ & $\beta^k$  &	 $\alpha^w$ & $\beta^w$\\
		[2pt]
		\hline
		%\noalign{\smallskip}
		1 &	MPEG-2 & 4.2$\pm$0.2 &	0.12$\pm$0.02	& 10$\pm$2 &	0.06$\pm$0.01 &	5$\pm$1 \\
		2 & DivX  & 4.7$\pm$0.2 &	0.22$\pm$0.05	& 13$\pm$5 &	0.12$\pm$0.05 &	8$\pm$3 \\ 
		3 & WMV9  & 4.7$\pm$0.2 &	0.32$\pm$0.1	& 15$\pm$5 &	0.22$\pm$0.08 &	10$\pm$3 \\
		[2pt]
		\hline
		\end{tabular}
	\end{center}
\end{table}

\begin{table}[!ht]
	\caption{\label{tablewimax}The values of the model coefficients for MPEG-2, MPEG-4 (DivX), and WMV9 codecs in a WiMAX network}
	\begin{center}
		\begin{tabular}{c@{\quad}c@{\quad}c@{\quad}c@{\quad}c@{\quad}c@{\quad}c}
		\hline\noalign{\smallskip}
		\textnumero & Codec &	$Q_{ideal}$ &	$\alpha^k$ & $\beta^k$  &	 $\alpha^w$ & $\beta^w$\\
		[2pt]
		\hline
		\noalign{\smallskip}
		1 &	MPEG-2 & 4.2$\pm$0.2 &	---	& --- &	0.2$\pm$0.1 &	15$\pm$0.5 \\
		2 & DivX  & 4.7$\pm$0.2 &	0.5$\pm$0.3	& 30$\pm$1 &	0.3$\pm$0.1 &	15$\pm$0.5 \\ 
		3 & WMV9  & 4.7$\pm$0.2 &	---	& --- &	0.3$\pm$0.1 &	15$\pm$0.5 \\
		[2pt]
		\hline
		\end{tabular}
	\end{center}
\end{table}

It should be noted that the packet losses $p$ in Tables~\ref{tablewifi}--\ref{tablewimax} of the coefficient values are measured in percentages rather than absolute proportions. Network jitter is expressed in seconds, not milliseconds.

All data are summarized in three tables. Each table describes the values for one type of wireless networking: Wi-Fi, 3G, or WiMAX, and the original video stream is encoded by different video codecs.

The data obtained as a result of the experiments were processed using the technique described in~\cite{sagatov}. All errors at the video level and network level were analysed and the subjective video quality $Q_{MOS}$ was found depending on the percentage packet loss $p$ and network jitter $j$. The coefficient values obtained are summarized in Table~\ref{tablewifi} for the Wi-Fi network, Table~\ref{table3g} for 3G, and Table~\ref{tablewimax} for mobile WiMAX.

On analysing the data we found that 80\% of the degradation in video quality was due to packet loss and only 20\% was due to network jitter (delay variation).

\section{Features of Wireless Networks}
General indicators of network quality on a good, acceptable, poor (GAP) scale~\cite{calyam} are summarized in Table~\ref{tablegap}.

\begin{table}
	\caption{\label{tablegap}Values of network quality on the GAP scale}
	\begin{center}
		\begin{tabular}{c@{\quad}>{\centering}m{2.7cm}@{\quad}>{\centering}m{3cm}@{\quad}>{\centering}m{3cm}@{\quad}>{\centering}m{1.5cm}}
		\hline\noalign{\smallskip}
		{\centering}\textnumero & Wireless Network &	Average percentage packet loss, \% &	Average network jitter, ms & Value on GAP scale \tabularnewline
		[2pt]
		\hline
		\noalign{\smallskip}
		1 &	Wi-Fi & ~~~~6~(poor) &	20 (acceptable)	& poor \tabularnewline
		2 & 3G  & $>$10~(poor) &	35 (acceptable)	& poor \tabularnewline
		3 & Mobile WiMAX  & ~~0.2~(good) &	15 (good)~~~~~~~~	& good \tabularnewline
		[2pt]
		\hline
		\end{tabular}
	\end{center}
\end{table}

Characteristics of WiMAX networks are comparable to those of Ethernet fixed networks. In experiments with a fixed WiMAX network the percentage packet loss is always close to 0\% and delay variation is of the order of 19~ms, even in tests conducted with a large competing traffic. According to~\cite{calyam}, this type of traffic is referred to as ``good" on the GAP scale.

The mobile WiMAX standard is characterized by a low percentage packet loss of 0.1--0.2\%. Jitter and connection bandwidth depend on the signal level in the modem, which the operator Yota describes on a four-point scale. When the device shows four points then the available bandwidth of incoming and outgoing channels is measured as several megabits per second and average network jitter is of the order of 4~ms. When degradation of the signal level to two points occurs, bandwidth is reduced to several hundred kilobits per second and average network jitter is of the order of 31~ms and increases strongly during the impairment of communication. According to~\cite{calyam}, reliable reception of this network traffic is referred to as ``good" on the GAP scale.

Qualitative characteristics of WiMAX networks are sufficiently high that even large values of the coefficients $\alpha$ and $\beta$ do not affect the quality of communication. Deterioration in the quality of the network on a MOS scale hovering around \mbox{0.3--0.4} and exceeding the estimate is good. At the same time smaller values of the coefficients for the Wi-Fi and 3G networks, taking into account the low quality of networks, show a significant (up to 2--2.5~points) deterioration in the video quality (see data in Tables~\ref{tablewifi}--~\ref{tablegap}). It should be noted that in practice there is a deviation from linear dependence; the coefficients $\alpha$ and $\beta$ for small values of $p$ and $j$ are significantly greater than for typical values in Wi-Fi and 3G networks.

The WiMAX network showed significantly better characteristics compared to Wi-Fi and 3G and is more suitable for real-time video transmission. Wi-Fi technology is suitable for real time video transmission only in the immediate vicinity of the access point. Video quality deteriorates significantly when the device moves a few tens of meters away. 3G networks have the worst results and the client is likely to be dissatisfied with the quality of the video without revision of hardware and software.

\section{Packet Duplication}
Earlier in~\cite{sagatov}, ways of achieving a significant increase in the quality of video transmission along a wireless network were formulated:

\begin{enumerate}
	\item Upgrade the video player on the receiving side in order to automatically reject duplicated RTP packets;
	\item The streaming server must duplicate packets containing information on key frames.
\end{enumerate}

In addition it was found that in 3G and WiMAX networks the packages are lost uniformly, but in the Wi-Fi networks packets are lost in groups at randomly distributed time intervals. Some of the equipment in 3G networks independently duplicates packets in RTP streams sent from the base station. This approach leads to a significant deterioration in the quality of communication using RTP/UDP protocols.

In the process of testing ideas to improve the video quality we have developed tools that

\begin{enumerate}
	\item drop duplicate packets at the receiving end;
	\item can duplicate packets that contain all the video frames or key frames only.
\end{enumerate}

This tool has been implemented in the Windows operating system using the Windows Driver Development Kit.

\begin{figure}
\centering
\includegraphics[width=12.2cm]{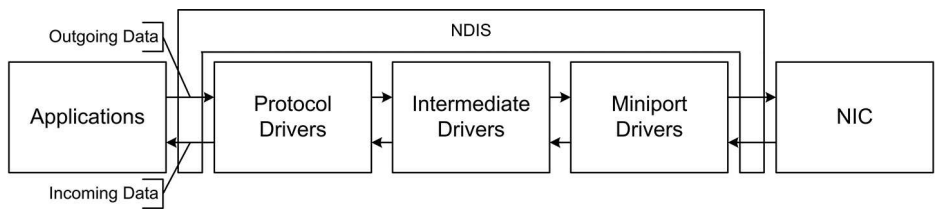}
\caption{Windows network drivers\textquoteright{} hierarchy}
\label{ndis}
\end{figure}

The Windows network drivers’ hierarchy~\cite{msdn} is presented in Figure~\ref{ndis}.

The tool which was developed is embedded in a chain of intermediate filters and receives all outgoing packets from the application after the processing of protocol drivers but before the transfer to miniport drivers. Such a position allows all outgoing packets to be duplicated, with knowledge of their type, size, and other service information with which they will be transferred to the network.~Also, the tool will be able to allocate RTP packets and key frames in them from the total number of outgoing packets. 

\begin{figure}[!ht]
\centering
\includegraphics[width=12.2cm]{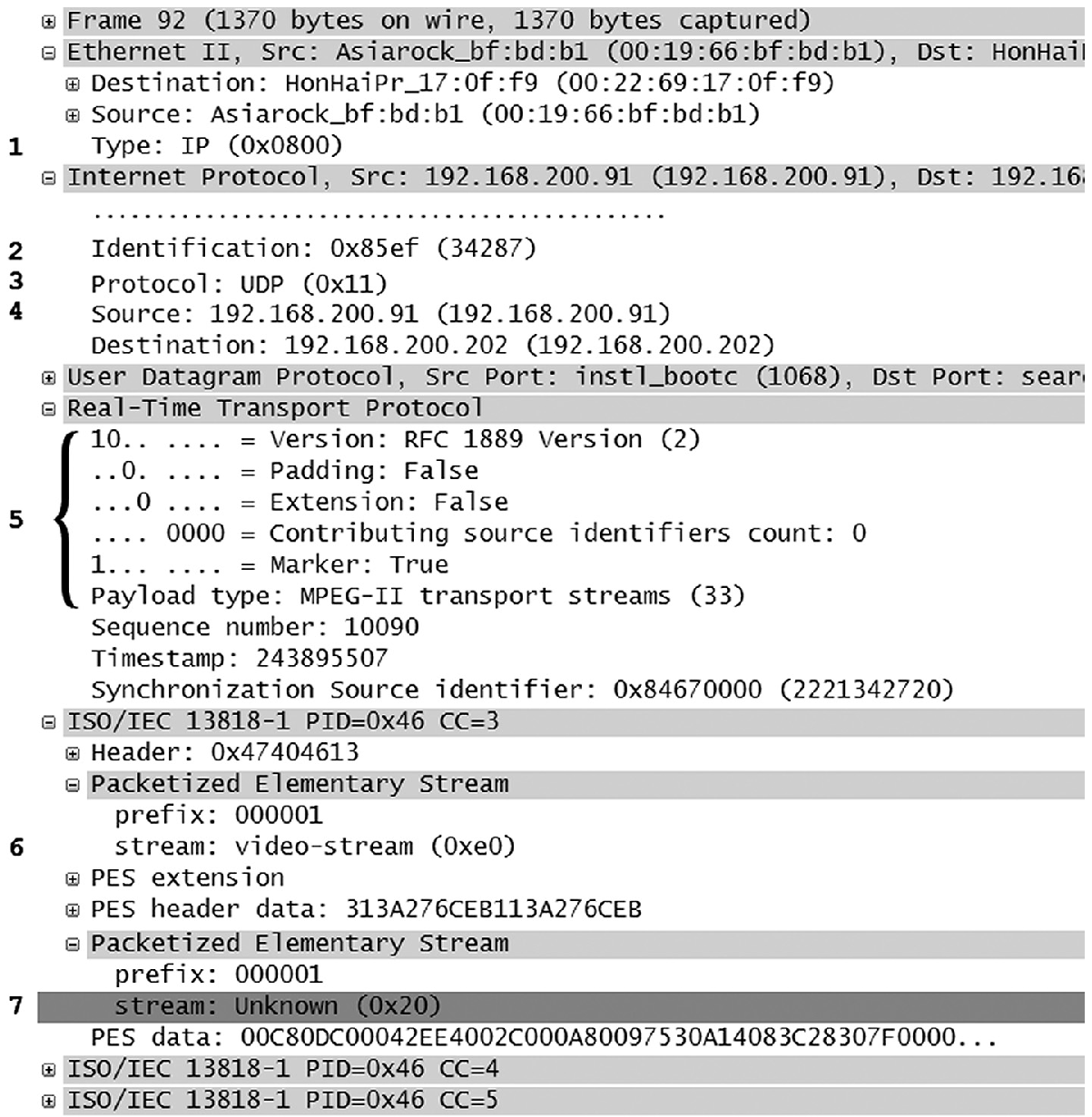}
\caption{Identification of key frames using WireShark}
\label{wireshark}
\end{figure}

Figure~\ref{wireshark} shows the scheme for determining the start of the key frame. The tool analyses fields 1 and 3 and determines the required type of packet (UDP) and whether the package is the fourth IP version. Then the existence and correctness of the RTP header (field 5) are checked. Key frames are characterized by certain data in fields 6 and 7, which are different for MPEG-2, MPEG-4 (DivX), and WMV9 codecs. If the data format indicates that it is a key frame then the packets with this frame can be duplicated.

\section{Effect of Duplication of Key Frames}
\begin{figure}
\centering
\includegraphics[width=12.2cm]{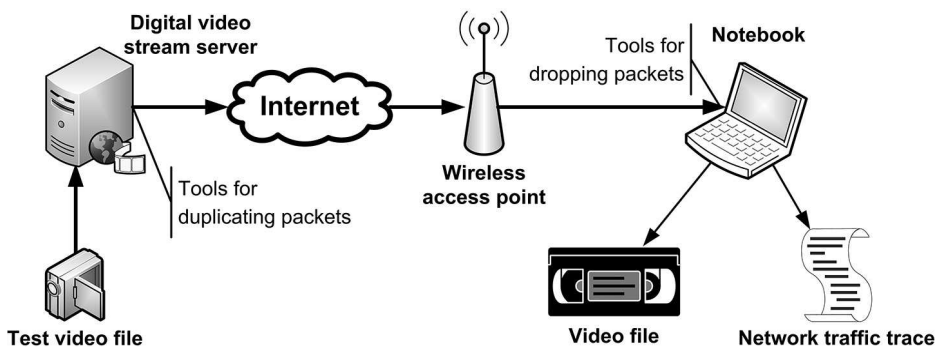}
\caption{Plan of the experiment with duplicate packets}
\label{plan}
\end{figure}

In order to test the hypothesis that the duplication of the key frames will lead to significant improvements in video quality we carried out a series of experiments (Figure~\ref{plan}). The tool described in the previous section allows frames to be duplicated on the transmitting side and dropped on the receiving side. This software has been installed in both the server and client in a wireless local area network of Wi-Fi (IEEE 802.11g) standard. Every video fragment encoded by MPEG-2, MPEG-4 (DivX), or WMV9 codecs was transferred through a local network three times: once without duplication, the second time with duplication of key frames only, and the third time with duplication of all frames. On the receiving side duplicated frames were dropped and the video was recorded to a file for analysis of the quality on the MOS scale.

It should be noted that the latest version of VideoLan VLC (January 2011) drops duplicate packets independently. Upon receipt of an RTP packet with sequence number $N$, VLC will automatically drop all subsequent packets with sequence numbers which do not exceed $N$. Violation of the order of packets due to network jitter leads to loss of all unordered RTP packets. Therefore the tool drops only those packets whose numbers coincide with those obtained previously.

Data on degradation of video quality $_{\Delta}Q$ in Wi-Fi networks for different types of duplication are presented in Table.

\begin{table}
	\caption{\label{tabledeg}The degradation of video quality under different types of packets duplication for a Wi-Fi network}
	\begin{center}
		\begin{tabular}{c@{\quad}>{\centering}m{1.5cm}@{\quad}>{\centering}m{2.5cm}@{\quad}>{\centering}m{2.5cm}@{\quad}>{\centering}m{2.5cm}}
		\hline\noalign{\smallskip}
		{\centering}\textnumero & Codec &	Without packets duplication &	With key frames duplication & With full packets duplication \tabularnewline
		[2pt]
		\hline
		\noalign{\smallskip}
		1 &	MPEG-2 & 0.7 & 0.3	& 1.0 \tabularnewline
		2 & DivX   & 1.2 & 0.4	& 1.5 \tabularnewline
		3 & WMV9   & 1.2 & 0.4	& 1.5 \tabularnewline
		[2pt]
		\hline
		\end{tabular}
	\end{center}
\end{table}

Duplicating key frames increases the volume of information transmitted by 7\% while the video quality is improved by a factor of almost three. Low quality levels occur when there is full duplication of video frames, due to the increased amount of information that is transmitted twice. The greater the streaming rate, the greater the percentage packet loss. This dependence is not linear and the percentage packet loss increases faster than the growth in the average network utilization. In addition the effect of breaking the packets order also increases with decreasing intervals between packets in the video stream.

\section{Conclusions}
This paper discusses ways to improve the quality of video streaming in wireless networks through packets duplication. In order to verify the hypothesis made earlier, a tool has been developed that allows key frames and all frames in the RTP stream to be duplicated. The article gives a detailed description of the functioning of this tool and an algorithm for determining key frames.

In experiments with the developed tool it was found that the duplication of key frames only is an ideal way to improve the quality of video streams. When traffic increases by 7\%, the quality of the received video is improved by a factor of almost three. Paradoxically, however, duplication of all the frames in the video stream leads to deterioration in the video quality even compared with the case without duplication.

In this study, an analysis was also conducted of a WMV codec and a WiMAX network, which was previously missing because of a lack of tools. New calculations based on additional data allowed the coefficient values of the analytical model to be specified and their final values to be presented. The most promising networks for video broadcasts should recognize fourth generation wireless networks, in particular WiMAX, whose quality is comparable to the quality of fixed networks.

\section{Introduction}

\end{document}